\documentclass[aps,prb,twocolumn,superscriptaddress,floatfix]{revtex4-2}
\usepackage{amsmath,amssymb,graphicx}
\usepackage{float}
\usepackage{placeins}
\usepackage{textgreek}

\begin{document}

\title{Second Harmonic Generation Spectroscopy of the Surface Charge Density Wave in the Weyl Semimetal CoSi}

\author{Awadhesh K. Das}
\affiliation{Department of Physics, Temple University, Philadelphia, PA 19122}

\author{Wesley E. Deeg}
\affiliation{Department of Physics, Temple University, Philadelphia, PA 19122}

\author{Sujan Subedi}
\affiliation{Department of Physics, Temple University, Philadelphia, PA 19122}

\author{Chandra Shekhar}
\affiliation{Max Planck Institute for Chemical Physics of Solids, 01187 Dresden, Germany}

\author{Claudia Felser}
\affiliation{Max Planck Institute for Chemical Physics of Solids, 01187 Dresden, Germany}

\author{Darius H. Torchinsky}
\email{dtorchin@temple.edu}
\affiliation{Department of Physics, Temple University, Philadelphia, PA 19122}

\date{\today}
%%%%%%%%%%%%%%%%%%%%%%%%%%%%%%%%%%%%%%%%%%%%%%%%%%%%%%%%%%%
\begin{abstract}
%%%%%%%%%%%%%%%%%%%%%%%%%%%%%%%%%%%%%%%%%%%%%%%%%%%%%%%%%%%
Using temperature-dependent rotational anisotropy second harmonic generation (RA-SHG), we identify a charge density wave (CDW) instability on the CoSi (001) face with an onset temperature at $90.0\pm0.8$ K. The SHG response tracks the order parameter amplitude, dominated by two nonlinear tensor elements, whose background-subtracted intensity evolves with temperature as a power law. The extracted critical exponent $\beta = 0.30 \pm 0.03$ is consistent with the 3D XY universality class, an unexpected result for a nominally two-dimensional surface layer. No corresponding anomaly is observed in the bulk response, establishing the transition as a purely surface-driven phase transition. We attribute this unexpected scaling to the coupling between surface Fermi-arc states and the bulk topology they are tied to, grounded in a previously observed intra-unit-cell phase relationship between surface sublayers that gives the order parameter intrinsic three-dimensional structure.
\end{abstract}
\maketitle
%%%%%%%%%%%%%%%%%%%%%%%%%%%%%%%%%%%%%%%%%%%%%%%%%%%%%%%%%%%
The surface of topological matter is inextricably linked to the bulk. In Weyl semimetals, this link is rooted in the bulk band structure, which hosts pairs of Weyl nodes with a conserved topological charge that cannot terminate at the surface. As a result, the bulk-boundary correspondence  dictates that the surface electronic structure comprise open, two-dimensional contours in momentum space, known as Fermi arcs \citep{armitage2018weyl}, that connect the nodes' projections across the surface \citep{wan2011topological, rao2019observation, sanchez2019topological}. The arcs are therefore as defining a feature of the Weyl phase as the quantized charge carried by the nodes themselves \citep{xu2015discovery}.

While band topology on its own produces the novel physics associated with protected surface states and quantized responses, the addition of strong electronic correlations allows symmetry-breaking order parameters to couple to topological invariants in ways that have no conventional analog, producing entirely new phases of matter and motivating a growing search for correlated topological systems \citep{rachel2018interacting}. Perhaps the most famous example is the Majorana fermions that arise from the coupling of a topological insulator to an s-wave superconductor via the proximity effect \citep{fu2008superconducting}. In Weyl semimetals, a charge density wave that opens a gap at the bulk Weyl nodes can drive the material into an axion insulating phase, a genuinely new topological state arising from the combination of order and topology in which the CDW phason is identified with the axion field, coupling to the electromagnetic field through a topological $\theta \mathbf{E}\cdot\mathbf{B}$ term \citep{wang2013chiral}. What an analogous ordering instability produces when it instead forms on the boundary arcs, rather than on the bulk nodes, remains unknown since no established framework describes the consequence of correlated order acting on a Fermi surface of this topology.

Fermi-arc nesting has been proposed as one possible route toward such a state, potentially producing interaction-driven surface instabilities such as charge-density-wave order \citep{bobrow2020monopole, shi2021charge, burkov2016review}. Recent STM measurements on CoSi show that the (001) surface supports precisely such a unidirectional, incommensurate CDW whose wave vector is locked to the crystal's structural chirality, with no corresponding ordering on the (111) surface, providing the first experimental foothold on this open question \citep{li2022chirality, rao2023charge}. However, further details on the CDW's nature remain undetermined; in particular, its universality class, which would reveal important information on both the effective dimensionality of the ordering and its microscopic coupling, has yet to be established.

Second harmonic generation (SHG) provides a sensitive probe of such surface-confined effects. Electric-dipole SHG requires broken inversion symmetry, and is therefore allowed wherever it is absent, whether at a static interface or across a symmetry-lowering phase transition. This has made SHG a standard tool for detecting symmetry-breaking phase transitions across a wide range of correlated systems, including charge density wave order \citep{ShenNature1989, Lupke1999, Sipe1987, zhang2022second}, magnetic and hidden-order transitions \citep{ahn2024electric}, and structural phase transitions \citep{jin2020observation}. Its non-contact nature and sensitivity to the order parameter further make it well suited to studying surface instabilities in topological semimetals \citep{Lupke1999, zhang2022second}.

Here, we report on rotational anisotropy second harmonic generation (RA-SHG) measurements on the (001) face of CoSi single crystals that reveal a clear second-order phase transition we attribute to the previously observed incommensurate surface CDW \citep{li2022chirality, rao2023charge}. CoSi belongs to the Sohncke cubic space group 198 (B20 family) and has two symmetry-protected multifold band crossings at the $\Gamma$ point and R point of the Brillouin zone \citep{tang2017multiple, bradlyn2016beyond} that are connected by large surface Fermi arcs. Study of this incommensurate CDW on the (001) surface therefore offers a platform for probing the interplay between electronic order and boundary states protected by bulk topology.

Significantly, our data reveal that the temperature dependence of the surface SHG signal intensity follows a power law with critical exponent $\beta = 0.30 \pm 0.03$, the first such determination for this transition. This result is consistent with the 3D XY universality class, an unexpected result given the nominally two-dimensional geometry of the surface layer. We argue that this apparent contradiction is resolved by an intra-unit-cell phase relationship between surface sublayers, previously identified by STM, which gives the CDW order parameter intrinsic three-dimensional character.

The SHG response at electric-dipole order is given by 
\begin{equation}
P_i(2\omega) = \chi^{(2)}_{ijk}(2\omega; \omega, \omega) E_j(\omega) E_k(\omega),
\end{equation}
where $P_i(2\omega)$ is the radiating second-harmonic polarization, $E_j(\omega)$ and $E_k(\omega)$ are components of the incident electric field, and $\chi^{(2)}_{ijk}(2\omega; \omega, \omega)$ is the second-order nonlinear susceptibility tensor, which encodes the point-group symmetry of the medium through Neumann's principle \citep{birss1964symmetry}. The (001) surface of CoSi is properly described by the C$_1$ point group, so all surface components of $\chi^{(2)}_{ijk}$ are symmetry-allowed. At normal incidence, the incident fields are purely in-plane ($E_x$ and $E_y$), restricting the surface response to components with $j,k = x,y$. Since a dipole oscillating along the beam propagation direction does not radiate along that axis, only the transverse ($i = x,y$) components contribute to the detected signal, leaving six independent surface tensor elements, i.e., $\chi^{(2)}_{xxx}, \chi^{(2)}_{xxy}, \chi^{(2)}_{xyy}, \chi^{(2)}_{yxx}, \chi^{(2)}_{yxy}$ and $\chi^{(2)}_{yyy}$ as the only probed elements. The same restrictions eliminate the bulk contribution as the only bulk-allowed element in this space group is $\chi^{(2)}_{xyz}$ (and all permutations to which it is equal) \citep{lu2022second}, which requires one factor of $E_x$, $E_y$, and $E_z$. With no incident $E_z$ at normal incidence, the sole surviving bulk term $P_z \propto \chi^{(2)}_{zxy}E_xE_y$ does not radiate, isolating the surface response.

\begin{figure}
    \centering
    \includegraphics[width=0.99\linewidth]{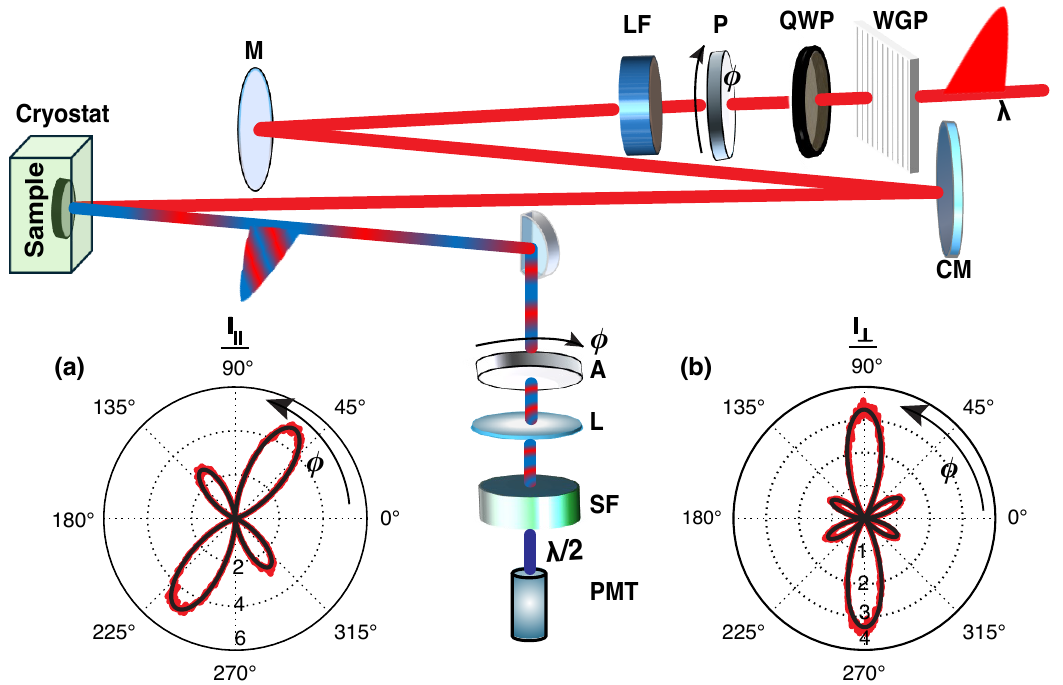}
    \caption{Schematic diagram of the experimental setup for SHG measurements. Optics are \textbf{WGP}-wire grid polarizer, \textbf{QWP}-quarter wave plate, \textbf{P}-polarizer, \textbf{LF}-longpass filter, \textbf{M}-mirror, \textbf{CM}-concave mirror, \textbf{A}-analyzer, \textbf{L}-lens, \textbf{SF}-shortpass filter and \textbf{PMT}-photomultiplier tube. The fundamental ($\omega$) and second harmonic ($2\omega$) beams are represented in red and blue, respectively. Polar plots (a) $I_{\parallel}$ and (b) $I_{\perp}$ show raw data (red) with fits (black) to surface point group C$_1$. Here \textbf{P} and \textbf{A} are co-rotated, i.e., parallel for $I_{\parallel}$ and perpendicular for $I_{\perp}$.}
    \label{fig:setup}
\end{figure}

Fig.~\ref{fig:setup} shows the experimental geometry used to perform rotational anisotropy SHG (RA-SHG) at near-normal incidence. A 1500~nm fundamental beam was generated by an optical parametric amplifier (OPA) pumped by a Ti:sapphire amplifier (5 kHz, $\sim$35 fs). The incident polarization was purified by a wire-grid polarizer and converted to circular polarization using a quarter-wave plate; a rotating linear polarizer mounted on a motorized rotation stage then set the final polarization angle $\phi$, with the intervening circular polarization step ensuring that the incident intensity remained constant throughout the rotation. Long-pass filters removed upstream SHG before the beam reached the sample, and the reflected SHG was spectrally filtered and detected with a photomultiplier tube.

RA-SHG measurements were performed on polished CoSi (001) single crystals (see Supplementary for details on crystal growth), mounted in a closed-cycle cryostat with a temperature stability of $\pm$ 0.3 K over the $50$--$115$ K range. X-ray diffraction confirmed the orientation of the cut and polished (001) surface to within $<1\%$. The incident beam was aligned along the crystallographic $c$-axis, with the $ab$-plane defining the transverse plane in which the polarization angle $\phi$ was varied.

\begin{figure}
 \centering
\includegraphics[width=0.9\linewidth]{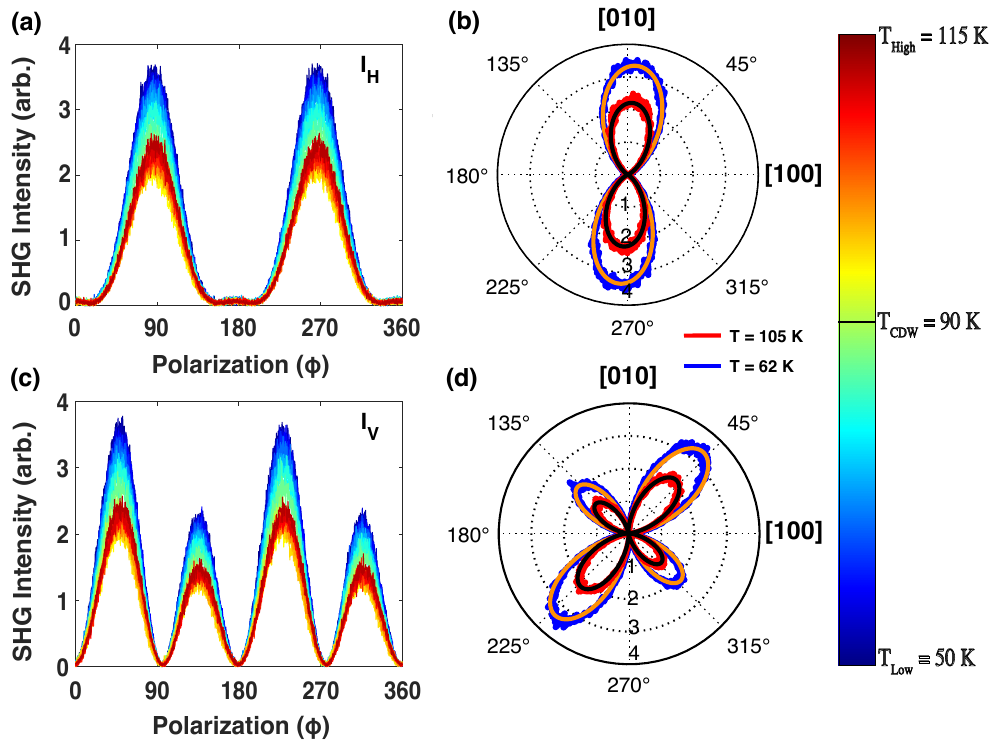}
\caption{RA-SHG measurements on the CoSi (001) surface showing the evolution of the angular response across the CDW transition. The SHG intensity as a function of input polarization angle $\phi$ is presented for the (\textbf{a}) fixed horizontal analyzer $I_H$, (\textbf{b}) fixed vertical analyzer $I_V$. Panels (\textbf{c}) and (\textbf{d}) show the corresponding polar plots at 105 K (red, $T>T_{CDW}$) and 62 K (blue, $T<T_{CDW}$), together with global fits to the surface symmetry point group (solid black and orange curves). A pronounced change in the angular anisotropy appears below $T_{CDW}$, signaling the onset of surface ordering.}
\label{fig:sur_raw}
\end{figure}

We measured the SHG intensity as a function of incident polarization angle $\phi$ in four geometries: (a) fixed horizontal analyzer ($I_H$), (b) fixed vertical analyzer ($I_V$), (c) co-rotating parallel ($I_{\parallel}$), (d) co-rotating perpendicular ($I_{\perp}$) (full geometry explained in Supplementary, S2). Rotating $\phi$ while recording the emitted SHG intensity at each geometry produces a polar plot whose angular shape and overall magnitude together encode the six surface tensor elements that can be recovered by a global fit. Fig.~\ref{fig:sur_raw} shows representative raw data for two of these geometries, $I_H$ and $I_V$, plotted on Cartesian axes as a function of $\phi$ across the measured temperature range, alongside polar plots of the same data at representative temperatures above and below $T_{CDW}$ (all four-geometry data, models, and fitting procedures are in Supplementary S3). The shape of these traces is identical at every temperature measured, while their overall size grows by nearly a factor of two below $T_{CDW}$.

This observation is to be expected. Because an incommensurate CDW's phase has no fixed relationship to the lattice, the transition does not change the C$_1$ symmetry, consistent with Landau theory's prohibition on an ordered phase acquiring higher symmetry than its parent. The allowed tensor elements are therefore the same across the transition, leaving the angular shape of the RA-SHG traces unchanged, with any signature of the CDW appearing as a change in the magnitude of these elements instead. Since $I(2\omega) \propto |P_i(2\omega)|^2 \propto |\sum_{jk}\chi^{(2)}_{ijk} E_j(\omega)E_k(\omega)|^2$, this magnitude change scales as the square of the CDW order parameter, as we show below, giving access to the critical exponent.

The coupling between the surface CDW order parameter and the SHG response can be understood within a perturbative treatment of the nonlinear response, where $\chi^{(2)}_{ijk}$ takes the general form
\begin{equation}\label{eq:SHG}
\begin{aligned}
     \chi_{ijk}^{(2)} \propto \sum_g \rho_{gg} \sum_{m,n}
\left[\frac{\langle g | \mu_i | n \rangle
\langle n | \mu_j | m \rangle
\langle m| \mu_k | g \rangle}{(\omega_{ng} - 2\omega)(\omega_{mg} - \omega)}\right]
\end{aligned}
\end{equation}
where $\rho_{gg}$ is the ground-state population factor, $g$ is the ground state, and $m$ and $n$ are the excited states \citep{armstrong1962interactions, boyd2008nonlinear, butcher1990elements, shen1984principles, hanna1979nonlinear}. The operators $\mu_n$ ($n = i, j, k$) are the electric dipole moment matrix elements, and $\omega_{ng}$, $\omega_{mg}$ are the corresponding resonance frequencies. The temperature dependence enters through both $\rho_{gg}$ and the matrix elements of Eq.~\eqref{eq:SHG}. As the charge-density-wave gap opens, the population of states near the Fermi level is redistributed, changing the ground-state electronic occupation and modifying the transition amplitudes that contribute to $\chi^{(2)}_{ijk}$. Because the surface point group remains C$_1$ throughout the transition, no symmetry constraint forbids a term in $\chi^{(2)}_{ijk}$ that varies linearly with the CDW order parameter. Since both factors retain finite normal-state values, their leading corrections combine linearly in $\Delta$, giving $I(2\omega) \propto \Delta^2$ as used above, and providing the rationale for a power-law fit.

To isolate the CDW-driven contribution to the nonlinear susceptibility, the background was removed at the tensor-element level before performing the power-law analysis. A high-temperature reference value was subtracted from each fitted tensor element, yielding the CDW-sensitive amplitude
\begin{equation*}
\Delta \chi_{ijk}^{(2)}(T) = \chi_{ijk}^{(2)}(T) - \chi_{ijk}^{(2)}(T > T_{CDW}).
\end{equation*}
The SHG intensities were then reconstructed from these background-subtracted tensor elements, $I_{2\omega}(\phi,T) = |\sum_{jk} \Delta\chi^{(2)}_{ijk}(T) E_j(\phi)E_k(\phi)|^2$, ensuring that the resulting curves reflect only the temperature-dependent CDW response. Because the SHG intensity is quadratic in $\chi^{(2)}_{ijk}$, subtracting the background directly from the intensity itself mixes background and CDW terms and distorts the temperature dependence; subtracting at the level of the tensor-elements instead preserves the linear coupling between $\chi^{(2)}_{ijk}$ and the order parameter established above. The temperature dependence of the reconstructed intensity was then fit to the power-law form $I_{2\omega} \propto (1-T/T_{C})^{2\beta}$.

\begin{figure}
    \centering
     \includegraphics[width=0.95\linewidth]{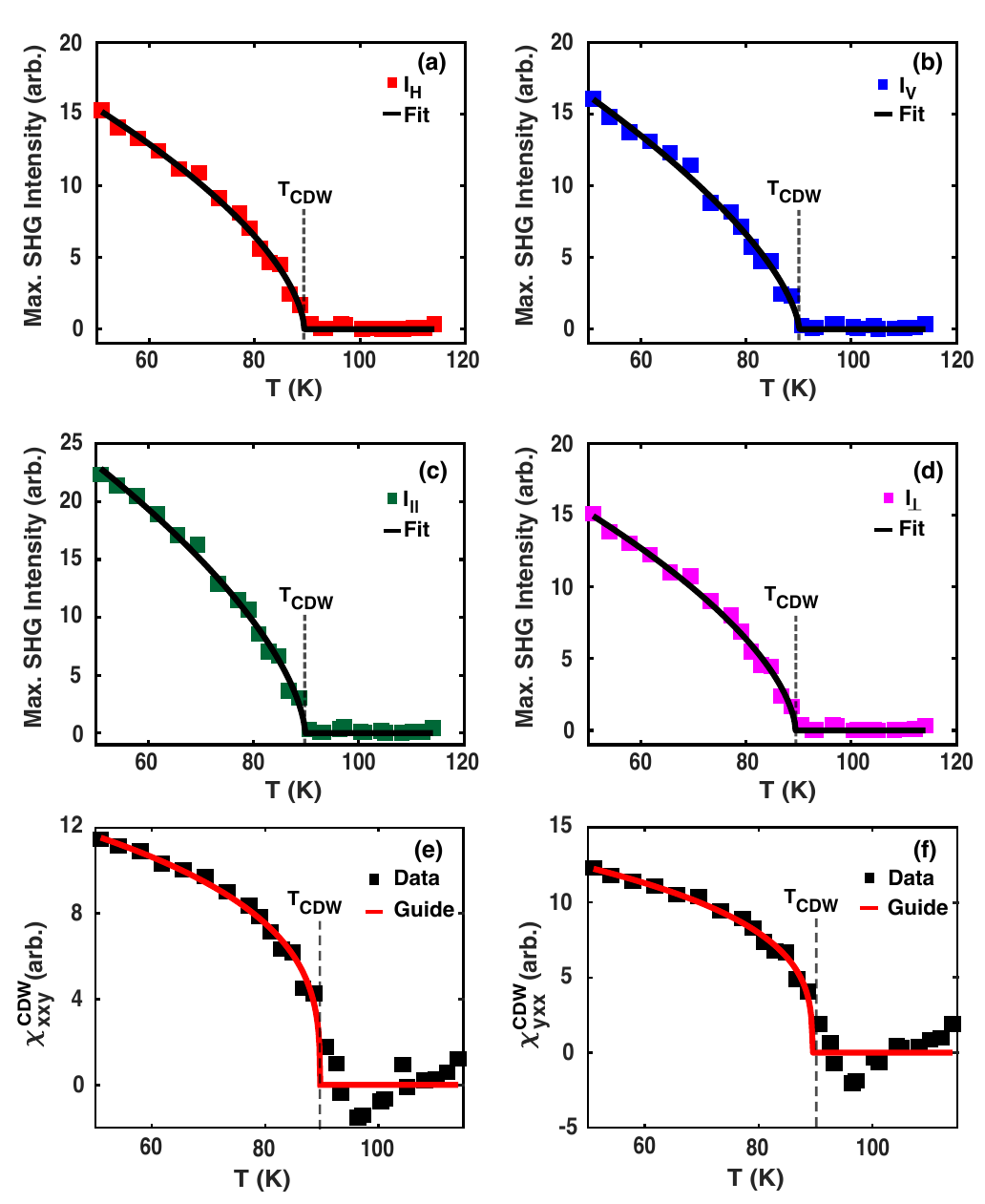}
    \caption{Panels \textbf{(a)}, \textbf{(b)}, \textbf{(c)}, and \textbf{(d)} show the background-subtracted temperature-dependent peak SHG intensity for the $I_H$, $I_V$, $I_{\parallel}$, and $I_{\perp}$ geometries, respectively, reconstructed using all six surface tensor elements (Supplementary Fig. 3) and evaluated through the expression in Supplementary Section S3. The transition temperature $T_{CDW} \sim 90.0$ K from the fit is marked by the black dashed lines. Panels \textbf{(e)} and \textbf{(f)} show the temperature dependence of the two dominant surface tensor elements, $\chi^{(2)}_{xxy}$ and $\chi^{(2)}_{yxx}$, extracted from global fits along with a line to guide the eye.}
    \label{fig:max_int}
\end{figure}

We extracted the maximum SHG intensity from each RA-SHG trace in all four polarization geometries ($I_H$, $I_V$, $I_{\parallel}$ and $I_{\perp}$). At each temperature, the raw traces were fit globally to recover the six surface tensor elements $\chi_{ijk}^{(2)}$ (Supplementary S3). These fits were tightly constrained, exhibiting low cross-parameter correlation and narrow confidence intervals at every temperature, indicating that the parameters obtained were not the result of compensating trade-offs between the six tensor elements. Fig.~\ref{fig:max_int} shows the background-subtracted peak intensities reconstructed using all six symmetry-allowed surface tensor elements (Supplementary Fig. 3). These SHG intensities remain nearly constant at high temperatures and develop clear features of a second-order phase transition near $\approx 90.0$ K.

To assess whether the observed transition extends beyond the surface, RA-SHG measurements were performed on the same CoSi (001) crystal with the incident beam at an off-normal angle of $\sim 30^\circ$, which introduces a $z$-component of the electric field and thereby probes the bulk directly. Because the mixed surface and bulk contributions at this incidence angle preclude the clean tensor decomposition used for the (001) and (111) geometries, we track the maximum SHG intensity rather than individual tensor elements. In contrast to the nearly twofold enhancement observed in the normal-incidence, surface-sensitive measurements, the off-normal data show only a weak temperature dependence across the transition (see Supplementary S4.2). The absence of a comparable enhancement in this geometry suggests that the observed CDW is predominantly confined to the (001) surface. Isolating the bulk response, however, requires a geometry that avoids the off-normal-incidence mixing of surface and bulk contributions.

\begin{figure}
    \centering
    \includegraphics[width=0.88\linewidth]{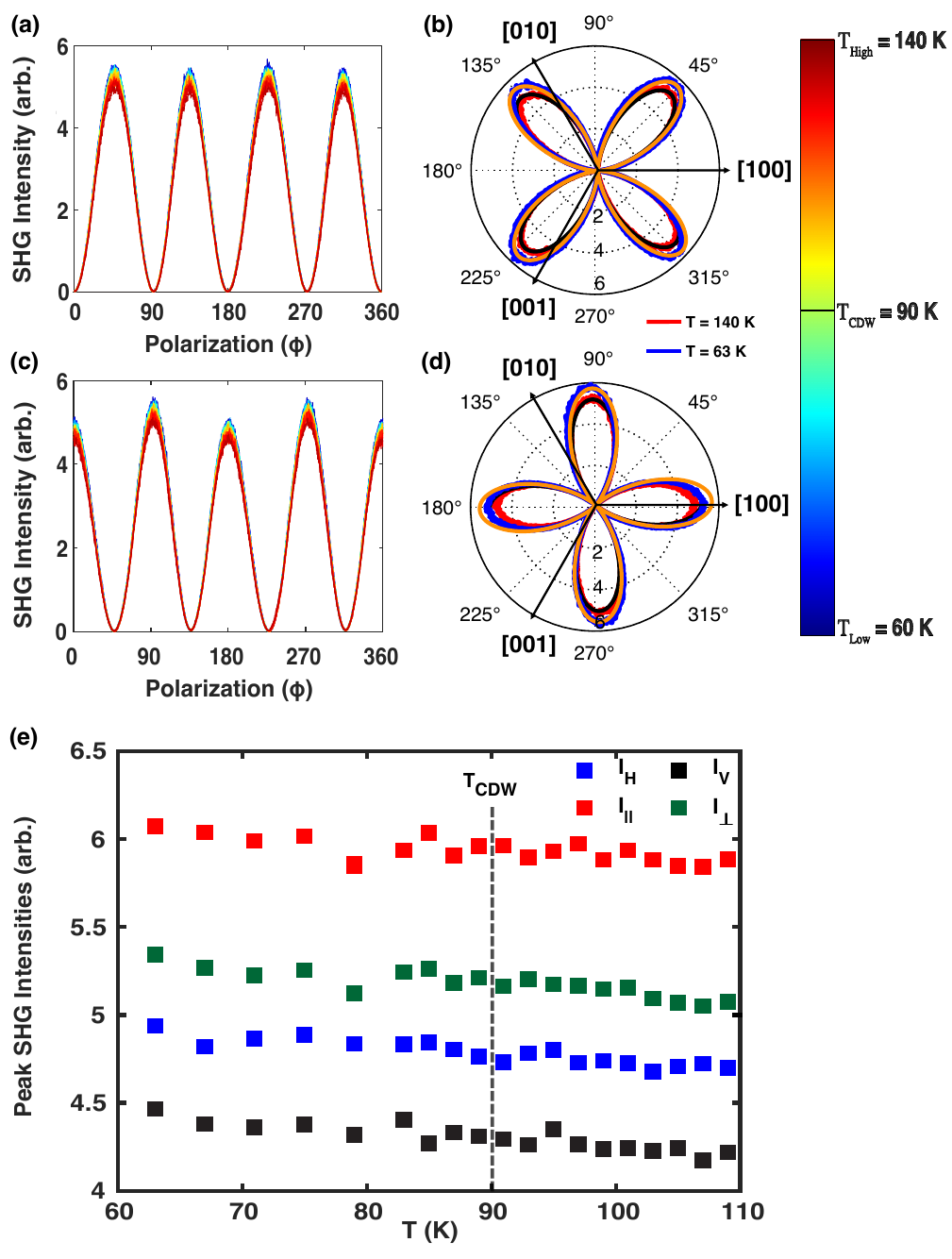}
    \caption{Temperature-dependent SHG response of the bulk CoSi over the temperature interval $60$--$110$ K. The SHG intensities for the $I_H$ and $I_V$ channels are shown in panels (\textbf{a}) and (\textbf{c}). The angular plots in panels (\textbf{b}) and (\textbf{d}) are identical above and below $T_{CDW}$ with global fits to Space Group 198, indicating the absence of temperature-driven modification on this face. Panel (\textbf{e}) shows the peak SHG intensities for $I_H$, $I_V$, $I_{\parallel}$ and $I_{\perp}$, offset by $\pm0.5$ for clarity. No significant temperature-driven transition is observed in the bulk SHG response.}
    \label{fig:bulk}
\end{figure}

Thus, we examined a separate crystal polished in the (111) orientation, following the approach used in prior SHG and photogalvanic studies of this space group \citep{deeg2025enantiomer, lu2022second, rees2020helicity, Ni2020Linear, Ni2021Giant}. Because $\chi^{(2)}_{xyz}$ is the sole bulk-allowed tensor element in this space group (established above), the (111) face provides a clean probe for any bulk contribution of the CDW order parameter. Fig.~\ref{fig:bulk}(a-d) shows the RA-SHG intensity measured as a function of incoming polarization ($\phi$) on the CoSi (111) surface over the temperature range 60--110 K at near-normal incidence. Measurements were performed in all four polarization geometries ($I_H$, $I_V$, $I_{\parallel}$, and $I_{\perp}$), but only the $I_H$ and $I_V$ channels are plotted for clarity; the full set of polarization channels is presented in the Supplementary Information (S4). The raw RA-SHG data and the corresponding fits to the model expressions (Supplementary, S4) above and below $T_{CDW}$ show that the SHG traces retain the same overall shape across the entire temperature range, with only a small, gradual increase in amplitude with decreasing temperature. In Fig.~\ref{fig:bulk}(e), the peak SHG intensities are plotted with a constant offset of $\pm0.5$ for clarity and show no evidence of a phase transition near $T_{CDW}$. The absence of any anomaly confirms that the CDW does not originate from the bulk and is confined to the (001) surface alone.

Fitting the temperature dependence of the reconstructed intensity to a power-law form $I_{2\omega} \propto (1-\frac{T}{T_{C}})^{2\beta}$ yields $T_{CDW} \sim 90.0\pm 0.8$ K and a critical exponent $\beta = 0.30 \pm 0.03$. This transition temperature is consistent with the surface CDW previously identified on CoSi (001) \citep{li2022chirality, rao2023charge}. Because SHG intensity scales as $|\chi^{(2)}_{ijk}|^2$ \citep{tuan2024symmetry, feng2012order, murphy2003surface, birss1964symmetry}, the extracted exponent corresponds to the order-parameter exponent, assuming the linear scaling between $\chi^{(2)}_{ijk}$ and the CDW order parameter justified above.

The CDW's incommensurability makes its order parameter complex, with the phase corresponding to a continuous translation of the modulation relative to the lattice, placing the transition naturally within the XY (U(1)) universality family rather than a discrete (Ising) or higher-symmetry (Heisenberg) class.  The power-law fit holds across the full measured range with no sign of the crossover expected near a 2D BKT transition. This behavior is inconsistent with the infinite-order transition itself, which lacks both a true order parameter and a conventional critical exponent, and the extracted value is likewise far from the mean-field limit ($\beta = 0.5$). Rather, our value is consistent with the 3D XY prediction, $\beta_{3DXY} = 0.3485(2)$ \citep{campostrini2001critical}, leaving 3D XY as the best-supported assignment.

The dominant tensor elements, $\chi^{(2)}_{yxx}$ and $\chi^{(2)}_{xxy}$, exhibit the strongest temperature dependence as seen in Fig.~\ref{fig:max_int}(e) and (f), while the remaining elements remain nearly an order of magnitude smaller and show no clear power-law scaling (Supplementary Fig. 3). The same two components dominate the surface response reported in RhSi \citep{rees2021direct}. This pattern contrasts sharply with PdGa, another member of this B20 family, whose photogalvanic tensor elements, which share the same symmetry-allowed structure as $\chi^{(2)}_{ijk}$, are instead all of comparable strength \citep{deeg2025enantiomer}. Unlike CoSi and RhSi, whose multifold band crossings sit close to the Fermi level, PdGa's Fermi level lies well above its multifold points, by roughly 0.5 and 1 eV at $\Gamma$ and R, respectively \citep{deeg2025enantiomer}, suppressing their contribution to the nonlinear response. The correspondence between a small number of dominant tensor elements and proximity of the multifold points to the Fermi level, common to both CoSi and RhSi despite their distinct microscopic origins, suggests a possible topological origin for this selectivity; resolving that connection will require further theoretical and experimental work.

\begin{figure}
    \centering
    \includegraphics[width=1\linewidth]{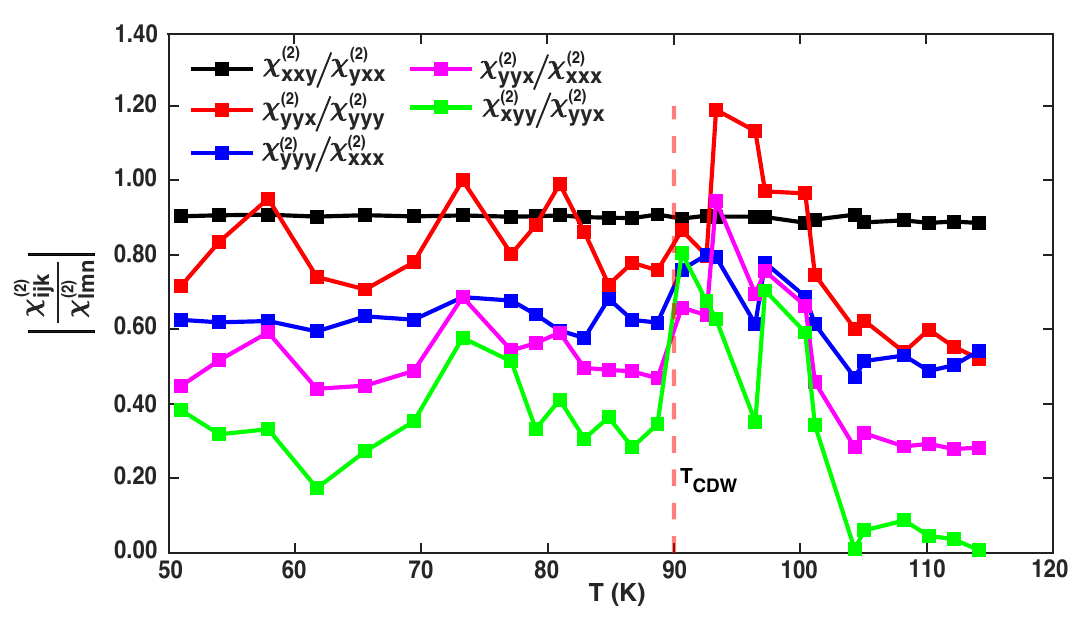}
    \caption{Temperature dependence of ratios among the six surface tensor elements. The ratio of the two dominant components, $\chi^{(2)}_{xxy}/\chi^{(2)}_{yxx}$ (black), remains constant to within a few percent across the entire measured range, including through $T_{CDW}$, indicating that both components track a common underlying quantity. Ratios among the four weak components ($\chi^{(2)}_{yyx}/\chi^{(2)}_{yyy}$, $\chi^{(2)}_{yyy}/\chi^{(2)}_{xxx}$, $\chi^{(2)}_{yyx}/\chi^{(2)}_{xxx}$, and $\chi^{(2)}_{xyy}/\chi^{(2)}_{yyx}$) vary only modestly below $T_{CDW}$ but fluctuate more substantially above it.}
    \label{fig:ratios}
\end{figure}

The temperature dependence of ratios among the six fitted tensor elements, plotted in Fig.~\ref{fig:ratios}, offers an independent, model-free assessment of the fitted transition temperature. Below approximately 90~K the ratios among all the tensor elements vary only modestly with temperature; above 90~K the remaining four fluctuate substantially, by more than a factor of two in some cases, while $\chi^{(2)}_{yxx}$ and $\chi^{(2)}_{xxy}$ do not. This is consistent with a loss of long-range phase coherence above $T_{CDW}$, from a single, macroscopically coherent order-parameter phase below the transition to an ensemble of short-range-ordered patches with uncorrelated phases above it. A coherent probe such as SHG, which sums contributions across the beam spot before detection, is expected to show fluctuating, non-monotonic individual tensor elements under these conditions even when the reconstructed intensity itself remains smooth. Significantly, the onset of this fluctuating behavior coincides with $T_{CDW}$ obtained separately from the power-law fit above, providing independent corroboration of the transition temperature.

We further note that the ratio $\chi^{(2)}_{xxy}/\chi^{(2)}_{yxx}$ remains constant to within a few percent across the full measured temperature range, including through the transition, as seen in Fig.~\ref{fig:ratios}. Because this ratio is insensitive to any common multiplicative factor, this indicates that both tensor elements track a single common factor rather than evolving independently, consistent with the dependence of $\chi^{(2)}_{ijk}$ on the population factor and matrix elements described above, though  our data are insufficient to distinguish which term is responsible.

These fluctuations in the individual $\chi_{ijk}^{(2)}$ that are absent from $I(T)$ itself are why we extract the critical exponent from $I(T)$ rather than from any individual tensor element. Because the present fits constrain each $\chi^{(2)}_{ijk}$ as a real quantity at each temperature, though it may in fact be complex, extracting the critical exponent from an individual component such as $\chi^{(2)}_{yxx}(T)$ or $\chi^{(2)}_{xxy}(T)$ yields exponents that vary erratically with fitting window and assumed $T_{CDW}$, whereas $I(T)$ does not, since $I(2\omega)$ comprises a coherent sum over all symmetry-allowed components in which only the overall phase of the response, and not the relative phase between components, is lost upon detection. We therefore consider $I(T)$ the more robust quantity for extracting the critical exponent and use it throughout.

Charge ordering on the CoSi (001) surface is limited to the topmost atomic layer, implying that the system is two-dimensional. According to the Mermin--Wagner theorem, a strictly 2D system with a continuous order parameter cannot sustain true long-range order at finite temperature \citep{mermin1966absence, hohenberg1967existence}, and should instead undergo a Berezinskii--Kosterlitz--Thouless (BKT) transition. In such a transition, correlations below $T_{BKT}$ decay algebraically rather than saturating to a nonzero order parameter, and the correlation length diverges exponentially, rather than as a power law, when approached from above \citep{kosterlitz1973ordering, kosterlitz1974critical}. No such signatures appear in our data. Rather, the order parameter shows three-dimensional, power-law growth across the full measured temperature range, with an exponent consistent with the 3D XY universality class and no indication of an intervening two-dimensional regime. This stands in contrast to the surface CDW in NbSe$_3$, where Brun \textit{et al.} identified a genuine two-dimensional BKT regime, with three-dimensional behavior recovered only near the bulk transition temperature \citep{brun2010surface}. The emergence of long-range order and a 3D XY-type critical exponent on a nominally two-dimensional surface is therefore unexpected.

This apparent contradiction may be reconciled once the microscopic stacking of the surface modulation is taken into account. STM topographs of the CDW across step edges of differing height parity reveal a nearly $\pi$ phase shift between adjacent sublayers within the unit cell, indicating that the charge modulation is not purely 2D but instead extends coherently along the [001] direction \citep{li2022chirality}. The sublayers exhibiting this phase relationship are separated by only 80 and 131 pm within a single unit cell \citep{li2022chirality}, comparable to interatomic bond lengths, indicating that the coupling responsible for this phase locking is intrinsically strong. These interlayer correlations give the surface CDW enough phase stiffness to overcome fluctuation-driven suppression, allowing long-range order to form despite the constraints imposed by the Mermin--Wagner theorem, placing the transition outside the BKT regime, consistent with a similar interlayer-coupling-driven crossover reported in IrTe$_2$ \citep{kim2023dimensional}.

Further support for this picture comes from the termination-dependence of the phase shift itself. Different surface terminations are known to host distinct arc states, as established both theoretically and by ARPES measurements on the Weyl semimetal NbP, which show inequivalent Fermi-arc states on opposite terminations \citep{souma2016direct, liu2016evolution}. In CoSi, different terminations correspond to the different sublayers exhibiting this phase relationship. The robust, reproducible $\pi$ phase shift observed specifically across odd, termination-changing steps, and its absence across even, termination-preserving steps \citep{li2022chirality}, is therefore evidence of genuine phase coherence linking the arc states hosted by inequivalent terminations, rather than a static structural coincidence between adjacent atomic layers.

These observations align with the broader understanding that Fermi-arc surface states are not isolated 2D bands but are instead boundary manifestations of the bulk band topology \citep{armitage2018weyl}. We suggest that this is a qualitatively different notion of ``two-dimensional'' than the one Mermin-Wagner's theorem was meant to describe. A two-dimensional electron gas or an exfoliated monolayer CDW material is a self-contained system, whereas a topological arc state has no independent existence apart from the three-dimensional bulk whose topology it bounds, so applying the Mermin-Wagner argument to it imports an assumption of dynamical self-containment that need not hold. The clear power-law growth of the order parameter observed here, with no signature of an intervening BKT regime, indicates that whatever mechanism provides this coupling supplies sufficient phase stiffness to drive the surface CDW into the 3D XY universality class.

In conclusion, our RA-SHG measurements reveal a temperature-driven enhancement of the nonlinear optical response on the CoSi (001) surface that originates from changes in the surface susceptibility rather than any modification of the surface symmetry, consistent with a CDW transition previously identified on this surface. Reconstructing the background-subtracted SHG intensity from all six surface tensor elements, we identify a continuous onset of order at $T_{CDW} = 90.0 \pm 0.8$ K, corroborated independently by the onset of fluctuations in the fitted tensor ratios. The transition follows power-law scaling with exponent $\beta = 0.30 \pm 0.03$, consistent with the 3D XY universality class despite the nominally two-dimensional geometry of the surface layer. We attribute this behavior to the coupling between surface Fermi-arc states and the bulk topology they are tied to. This coupling is grounded in the intra-unit-cell $\pi$ phase relationship between sublayers established by STM, which gives the order parameter intrinsic three-dimensional structure. Rooted in the arcs' dependence on the bulk topological invariant, this coupling can supply the phase stiffness needed for 3D XY criticality, though its specific microscopic origin remains an open question.

Complementary measurements on the (111) face show no anomaly in the bulk-allowed tensor element $\chi_{xyz}^{(2)}$, confirming that the ordering is confined to the (001) surface. Our results suggest that the instability driving the ordering is electronic in nature and associated with surface Fermi arc states, providing a platform in which topology and electronic interactions combine to produce emergent order. More broadly, this demonstrates that topological surface states can host correlated electronic order as an intrinsic, surface-confined phenomenon, suggesting that similar instabilities may emerge in other Weyl semimetals with extended Fermi arcs.

%%%%%%%%%%%%%%%%%%%%%%%%%%%%%%%%%%%%%%%%%%%%%%%%%%%%%%%%%%%
\section*{Acknowledgements}
%%%%%%%%%%%%%%%%%%%%%%%%%%%%%%%%%%%%%%%%%%%%%%%%%%%%%%%%%%%
The work at Temple University was supported by the National Science Foundation under Grant No. NSF/DMR-1945222. The work was also financially supported by the Deutsche Forschungsgemeinschaft (DFG, German Research Foundation) through SFB 1143 (project ID 247310070), QUAST (project ID FOR 5249), the W\"urzburg-Dresden Cluster of Excellence on Complexity, Topology and Dynamics in Quantum Matter---ct.qmat (EXC 2147, project ID 390858490), and EXQIRAL (No. 101131579).

%%%%%%%%%%%%%%%%%%%%%%%%%%%%%%%%%%%%%%%%%%%%%%%%%%%%%%%%%%%
\bibliographystyle{unsrtnat}
\bibliography{references}

\end{document}